# SELF–GRAVITY AND DISSIPATION IN POLAR RINGS

John Dubinski[1] and Dimitris M. Christodoulou[1,2]

October 5, 1993

[1] Harvard–Smithsonian Center for Astrophysics, 60 Garden St., Cambridge, MA 02138

[2] Virginia Institute for Theoretical Astronomy, Department of Astronomy, University of Virginia, P.O. Box 3818, Charlottesville, VA 22903

# ABSTRACT

Studies of inclined rings inside galaxy potentials have mostly considered the influence of self–gravity and viscous dissipation separately. In this study, we construct models of highly-inclined ("polar") rings in an external potential including both self–gravity and dissipation due to a drag force. We do not include pressure forces and thus ignore shock heating that dominates the evolution of gaseous rings inside *strongly* nonspherical potentials. We adopt an oblate spheroidal scale–free logarithmic potential with axis ratio $q = 0.85$ and an initial inclination of $80°$ for the self-gravitating rings. We find that stellar (dissipationless) rings suffer from mass loss during their evolution. Mass loss also drives a secular change of the mean inclination toward the poles of the potential. As much as half of the ring mass escapes in the process and forms an inner and an outer shell of precessing orbits. If the remaining mass is more than $\sim 0.02$ of the enclosed galaxy mass, rings remain bound and do not fall apart from differential precession. The rings precess at a constant rate for more than a precession period $\tau_p$ finding the configuration predicted by Sparke in 1986 which warps at larger radii toward the poles of the potential. We model shear viscosity with a velocity-dependent drag force and find that nuclear inflow dominates over self–gravity if the characteristic viscous inflow time scale $\tau_{vi}$ is shorter than $\sim 25\tau_p$. Rings with $\tau_{vi}/\tau_p \lesssim 25$ collapse toward the nucleus of the potential within one precession period independent of the amount of self–gravity. Our results imply that stars and gas in real polar rings exhibit markedly different dynamical evolutions.

*Subject headings:* galaxies: ISM – galaxies: kinematics and dynamics – galaxies: structure – method: numerical

# 1 INTRODUCTION

Over the past fifteen years, various workers have observed polar-ring galaxies and determined their structure and kinematics (*e.g.*, Schechter & Gunn 1978; Schweizer, Whitmore, & Rubin 1983; Whitmore 1984; Whitmore, McElroy, & Schweizer 1987; van Gorkom, Schechter, & Kristian 1987; Whitmore *et al.* 1990). Most observed rings orbit almost at right angles to the planes of their central galaxies with virtually no objects at intermediate or low inclinations ($\lesssim 50°$) (e.g. Athanassoula & Bosma 1985; see also Tohline 1990 and Sackett 1991). The unusual orientation of the rings implies that polar–ring galaxies are probably the results of accretion of mass onto the central host galaxies either through the merger of a smaller, gas-rich galaxy or through the interaction with a passing galaxy (e.g. Whitmore *et al.* 1987). Their frequency in the universe can potentially tell us something about the merger/accretion rate of galaxies. Schweizer, Whitmore, & Rubin (1983) and Sackett & Sparke (1990) argued that substantial flattening of the dark halos surrounding polar–ring galaxies is necessary to explain why rings are observed only at high inclinations. Low-inclination accretion events never became rings because of the instability resulting from a large amount of differential precession induced by a flattened halo. In fact, the gross geometric shapes of dark halos can be determined by measuring the ratio of the ring and galaxy orbital velocities. The results of Whitmore, McElroy, & Schweizer (1987) suggest that the overall potential well is nearly spherical (potential axis ratios $0.95 \leq c/a \leq 1.05$) in two observed polar–ring galaxies and a little more flattened in NGC 4650A ($c/a = 0.86$ in the potential implying an axis ratio $Z/R \approx 0.60$ in the density). On the other hand, detailed model fits to NGC 4650A by Sackett & Sparke (1990) suggest that if the dark halo is dominant then it may be considerably more flattened ($Z/R$ as small as 0.2 in the density) and that the ring may be much more massive than previously reported. These results suggest that at least some dark halos are flattened although more spherical halos cannot be excluded. Differential precession can quickly destroy a ring so either polar rings are short-lived phenomena (Schweizer *et al.* 1983) or a stabilizing mechanism is required to counteract the destructive shear of differential precession. At least some of the well–studied polar rings are probably as old as the host galaxies (e.g. A0136-0801, Whitmore *et al.* 1987), so the issue of long-term stability of precessing rings has been the focus of many recent studies.

Several stability mechanisms for polar rings have been put forward including the settling of gaseous rings in preferred planes of triaxial potentials (Steiman-Cameron & Durisen 1982), self-gravity (Sparke 1986), and radiative cooling of the gas (Katz & Rix 1992; Christodoulou *et al.* 1992). Steiman-Cameron & Durisen (1982, 1984; see also a review by Steiman–Cameron 1991) argued that highly inclined differentially–precessing gaseous rings should settle into the polar preferred plane of a triaxial potential and smoothed–particle hydrodynamical (SPH) simulations (Habe & Ikeuchi 1988; Varnas 1990) demonstrated how rings settle into this plane. Sparke (1986) demonstrated that a model composed of gravitating spinning ringlets embedded in a flattened potential can find a stable equilibrium in which all ringlets precess together at a constant rate. Only self-gravitating rings with relatively high inclinations are stable. The equilibrium arrangement of ringlets warps towards the poles of the potential at larger radii. Katz and Rix (1992) simulated a gaseous polar ring using an SPH code and also found a stable constantly–precessing equilibrium. The ring warps *away* from the poles settling into a surface of constant precession inside the external potential. The addition of self-gravity to these simulations did not change the results. Rings survive only at high inclinations where precession frequencies are relatively small, gas–cloud collisions are minimized, and shock formation is avoided because of efficient cooling processes. Note that this mechanism cannot operate at low or moderate inclinations where differential precession frequencies for gas–particle orbits are relatively large and rings fall apart before they find the long–lived state. Also it does not operate inside prolate potentials (see Christodoulou *et al.* 1992). Although dissipation can allow gaseous rings to find quasi-stable precessing states at high inclinations, the stellar populations in real polar rings must be stabilized by a different mechanism. Self-gravity is such a stabilizing agent (Sparke 1986, 1991; Sparke & Casertano 1988; Casertano 1991).

In this paper, we examine the relative importance of self–gravity, external gravity, and viscous dissipation on the stability of polar rings evolving inside a logarithmic potential. We do not include gas dynamics (pressure effects and shock formation) but we incorporate shear viscosity through a velocity–dependent drag force to mimic viscous dissipation. We use initial conditions appropriate for observed polar rings: the external oblate spheroidal potential well is only moderately nonspherical with an axis ratio of $c/a = 0.85$ corresponding to $Z/R \approx 0.57$ in the density; the self–gravitating rings are highly inclined to the equatorial plane of the external potential ($i_o = 80°$) and rather extended with a full width to radius ratio of 0.4. In §2, we present analytical estimates of the critical mass ratio and the critical viscous time above which self–gravity cannot be neglected and viscosity does not play an important role, respectively. We also present orbit integrations of test–particles inside a combination of potentials: an external, scale–free logarithmic potential and a time–varying potential due to a constantly–precessing, highly inclined, self–gravitating ring. In §3, we present $N$–body simulations of rings with and without an additional drag force that mimics dissipation. These simulations confirm the analytically derived critical values and our results from orbit integrations. They also show some interesting features of the evolution of particle rings (extensive mass loss from the rings and a secular increase of the mean inclination toward the poles of the potential) that cannot be seen in the ringlet models of Sparke and collaborators (*e.g.*, Hofner & Sparke 1991). We discuss our results in §4 and summarize our conclusions in §5.

## 2 CRITICAL VALUES AND ORBIT INTEGRATIONS

### 2.1 Definitions

For the galactic potential, we adopt a rigid axisymmetric scale–free logarithmic potential (Habe & Ikeuchi 1985),

$$\Phi_{ext}(R, \phi, Z) = \frac{1}{2} v_o^2 \ln{(R^2 + \frac{Z^2}{q^2})}, \qquad (1)$$

where $(R, \phi, Z)$ are cylindrical coordinates, $q \equiv c/a$ is the axis ratio of the potential, and $v_o$ is the rotation velocity. Although the mass distribution of this flattened logarithmic potential is complicated, the mass enclosed within a reference radius $r_o = (R_o^2 + Z_o^2)^{1/2}$ is exactly

$$M_H = \frac{r_o v_o^2}{G}, \qquad (2)$$

*independent* of $q$ (use Gauss's law on a sphere).

A circular orbit of radius $r$ placed in this potential at an angle $i_o$ to the equatorial plane precesses with frequency (Steiman–Cameron & Durisen 1990)

$$\omega_p(r) = -\frac{3}{4}\frac{v_o}{r}\eta \cos i_o, \qquad (3)$$

where the quadrupole coefficient $\eta$ is a function of the axis ratio only, i.e.,

$$\eta \equiv \frac{2(1-q^2)}{1+2q^2}. \qquad (4)$$

This is a good approximation for $q > 0.8$ as shown by comparing to the precession rates of orbits measured from test–particle integrations. Typically, the above equations underestimate the true rate by 10% when $q = 0.8$ and the error decreases with increasing $q$.

We also consider an inclined self–gravitating ring. The ring is homogeneous with surface density $\sigma$ and extends between an inner $(r_-)$ and an outer $(r_+)$ radius. The reference radius $r_o$ can now be taken as the middle of the ring, i.e., $r_o \equiv (r_+ + r_-)/2$. The mass of the ring is

$$M_R = \pi(r_+^2 - r_-^2)\sigma. \qquad (5)$$

We define the full radial width of the ring

$$\Delta r \equiv r_+ - r_-, \qquad (6)$$

and the ring/galaxy mass ratio

$$\mu \equiv \frac{M_R}{M_H}. \qquad (7)$$

We also define a coefficient of kinematic viscosity, $\nu$, to characterize the dissipative effect of gaseous collisions.

We can now define four relevant time scales at radius $r_o$ which characterize the dynamics of the model. They are the orbital period

$$\tau_o \equiv \frac{2\pi r_o}{v_o}, \qquad (8)$$

the precession period

$$\tau_p \equiv \frac{2\pi}{|\omega_p(r_o)|}, \qquad (9)$$

the viscous inflow time

$$\tau_{vi} \equiv \frac{r_o^2}{\nu}, \qquad (10)$$

and the viscous spreading time

$$\tau_{vs} \equiv \frac{(\Delta r/2)^2}{\nu}. \tag{11}$$

Finally, we adopt three useful time scale ratios, i.e.,

$$T_p \equiv \frac{\tau_p}{\tau_o}, \tag{12}$$

$$T_{vi} \equiv \frac{\tau_{vi}}{\tau_o}, \tag{13}$$

and

$$T_{vs} \equiv \frac{\tau_{vs}}{\tau_o}, \tag{14}$$

which measure the number of orbital periods for the precession and viscous time scales. Generally, the orbital period is shorter than the precession period which in turn is shorter than the viscous time scales, i.e., $\tau_o < \tau_p \leq \tau_{vs} < \tau_{vi}$, or equivalently $1 < T_p \leq T_{vs} < T_{vi}$.

## 2.2 Analytical Estimates for Self–gravity

We estimate the critical mass ratio by using the largest possible magnitude of the force from the external gravity perpendicular to the ring, i.e.,

$$|F_\perp| = \epsilon \Big(\frac{v_o^2}{r_o}\Big) \frac{\sin i_o \cos i_o}{1 + \epsilon \sin^2 i_o}, \tag{15}$$

(Christodoulou *et al.* 1992) where we have assumed that $\epsilon \equiv q^{-2} - 1 > 0$. Comparing the magnitude of this force to an approximate expression for the local restoring force from self–gravity perpendicular to the plane of the ring, i.e., $F_{self} = 2\pi G \sigma$, and using equations (2) and (5)–(7), we find that

$$\mu_{crit} = \epsilon \Big(\frac{\Delta r}{r_o}\Big) \frac{\sin i_o \cos i_o}{1 + \epsilon \sin^2 i_o}. \tag{16}$$

In what follows, we adopt a set of standard model parameters representative of highly inclined, radially extended, self–gravitating rings inside a spheroidal potential: $i_o = 80°$; $q = 0.85$ implying that $\eta = 0.227$ and $T_p = 34$; and $r_+/r_o = 1.2$, $r_-/r_o = 0.8$ implying that $\Delta r/r_o = 0.4$. Using these values in equation (16), we find that

$$\mu_{crit} \approx 0.02. \tag{17}$$

## 2.3 Orbital Integrations with Self–gravity

The critical mass ratio $\mu_{crit}$ obtained analytically in §2.2 can be checked numerically by performing orbit integrations. Toward this end, we have developed an orbit integrator that combines both external gravity and self–gravity through the equations of §2.1. The computations are performed in a coordinate frame $(x, y, z)$ attached to the self–gravitating,

uniformly–precessing, homogeneous ring. The ring lies in the $xy$ plane of this frame while its symmetry axis $z$ precesses about the symmetry axis $Z$ of the spheroidal potential with a specified frequency $\omega_f$. The gravitational potential of the ring's self-gravity $\Phi_{self}$ is computed numerically by Toomre's (1962) method while the external potential $\Phi_{ext}$ is computed analytically after equation (1) is transformed to the ring's frame. If we assume that the external potential is inclined by an angle $i_o$ about the $x$ axis, the equation of motion for a test–particle in the precessing frame $(x, y, z)$ is (cf. Binney & Tremaine 1987)

$$\ddot{\mathbf{r}} = -\nabla(\Phi_{ext} + \Phi_{self}) - 2(\mathbf{\Omega_R} \times \dot{\mathbf{r}}) - \mathbf{\Omega_R} \times (\mathbf{\Omega_R} \times \mathbf{r}), \tag{18}$$

where dots denote time derivatives, $\mathbf{r} = (x, y, z)$, and $\mathbf{\Omega_R} = (0, -\omega_f \sin i_o, +\omega_f \cos i_o)$ is the precession frequency in the ring's frame. The initial velocity $\mathbf{v}$ of a test–particle is determined in the $(x, y, z)$ frame by the equation

$$\mathbf{v} = \mathbf{v_o} - \mathbf{\Omega_R} \times \mathbf{r}, \tag{19}$$

neglecting the ring's self-gravity.

We adopt the normalization $G = M_H = r_o = 1$ in which case $v_o = 1$ as well. We are then free to specify independently the following quantities: the axis ratio $q$ of the potential; the inclination $i_o$, the mass $M_R$, the precession frequency $\omega_f$, and the radial extent of the ring through the radii $r_-$ and $r_+$; and the location and velocity vector $\mathbf{v_o}$ of an orbiting test–particle. We use the following initial conditions: $q = 0.85$, $i_o = 80°$, $r_- = 0.8$, and $r_+ = 1.2$. The test–particle initially orbits in the plane of the ring at radius $r$ with azimuthal velocity $v_o$ while the ring precesses with the frequency predicted by equation (3) with $r = r_o$, i.e., $\omega_f = \omega_p(r_o)$. The mass ratio $\mu$ and the orbital radius $r$ are free parameters. The approximation $\omega_f = \omega_p(r_o)$ is justified by Figure 6 below.

If self–gravity is important, test–particles initially orbiting near the plane of the ring should remain bound to the ring. If self–gravity is not effective, test–particles should precess away from the plane of the ring. The orbit integrations are designed to demonstrate the effectiveness of a ring's self-gravity in locking surrounding orbits into coprecession and are not intended to represent self-consistent orbits in a self-gravitating ring. Clearly the subtleties of ring warping and internal redistribution of angular momentum involved in self-gravitating models are neglected. Nevertheless, the integrations provide a simple consistency check of the analytical prediction that a relatively small amount of self-gravity is sufficient to lock a ring into solid-body precession. Self-consistent $N$-body models are discussed in §3 below.

Figure 1 displays orbit integrations for $\mu = 0.03$ up to $t = 80\tau_o$. The panels on the left illustrate projections of an orbit with $r = r_-$ initially while $r = r_o$ initially for the panels on the right. Since the $xy$ plane is the symmetry plane of the precessing ring, we see that both orbits are locked into coprecession. In contrast, Figure 2 displays the corresponding orbit projections for a model with $\mu = 0.01$ up to $t = 80\tau_o$. Self–gravity is no longer sufficient to enforce coprecession for the adopted or any other value of $\omega_f$. All orbits computed with $\mu = 0.01$ and different combinations of $\omega_f$ and $r$ precess out of the plane of the ring and populate the surrounding volume.

We can delimit the region where test–particles are bound to the ring by integrating orbits from different initial radii and relative inclinations. Orbits initially close to $r = r_+$ do not coprecess for $\mu \lesssim 0.1$ but even these particles stay with the ring for many orbits before escaping if $0.02 \leq \mu \leq 0.1$. We have also examined orbits with $r = r_o$ initially inclined by up to $\sim 5°$ relative to the ring (or, equivalently, a height $\Delta z \approx 0.1$ above the ring plane). For $\mu = 0.01$, the test–particles precess away from the plane of the ring after a few orbits while for $\mu = 0.03$, the inclined orbits simply oscillate about the ring plane remaining bound for over a precession period.

In summary, by varying $\mu$ and $r$ independently, we have determined the critical mass ratio for which test–particles remain bound to the ring. For $\mu > \mu_{crit} \approx 0.02$ test–particles at a wide range of radii generally precess with the ring while for $\mu < \mu_{crit}$ all test–particles precess away within a few orbits. Despite the limitations inherent in these calculations, the derived value for $\mu_{crit}$ is in good agreement with the rough analytical estimates in §2.2.

### 2.4 Analytical Estimates for Viscosity

We determine the critical viscosity coefficient $\nu_{crit}$ above which viscosity competes with precession and cannot be neglected by using the shortest of the two viscous times, i.e., the viscous spreading time $\tau_{vs}$ defined by equation (11). The critical viscosity coefficient $\nu_{crit}$ is determined by setting $\tau_{vs} = \tau_p$, i.e.,

$$\nu_{crit} = \frac{(\Delta r/2)^2}{\tau_p}, \qquad (20)$$

where $\Delta r$ and $\tau_p$ are defined by equations (6) and (9), respectively. Using our set of standard model parameters (§2.2), we find that $T_{vi,crit} = 850$ and $\tau_{vi,crit} = 25\tau_p$. Viscous inflow can be driven by collisions between gas–clouds on adjacent tangential orbits in the ring (Steiman-Cameron & Durisen 1988). In a logarithmic potential, one can show that the shear viscous force is equivalent to a drag force

$$F_{drag} = -\nu \frac{v_\theta}{r^2}, \qquad (21)$$

where $v_\theta$ is the azimuthal velocity of each orbit. Assuming that $v_\theta$ is constant, an initially circular orbit spirals inward according to the equation

$$r^2 = r_o^2 - 2\nu t, \qquad (22)$$

where $r_o$ is the initial radius of the orbit and $t$ is the time. [This derivation is analogous to that of the effect of dynamical friction on globular clusters (Binney & Tremaine 1987)]. The time for an orbit to spiral into the center is $T_{vi}/2$ initial orbits. For $\nu = \nu_{crit}$ the inflow time is therefore $T_{vi,crit}/2 = 425$ orbits, generally much larger than a Hubble time. On the other hand, it takes the first 153 orbits (36% of the time) for the particle to cross the inner edge of the ring ($r = r_-$).

We can also compare the viscosity estimates of Steiman–Cameron & Durisen (1988) to the derived critical values. Using a cloud–fluid approach to the problem, these authors argue that viscosity in real galaxies is expected to be smaller than $\nu_{max} \sim 0.01 r_o^2/\tau_o$ (see their equations [49,51] and the following discussion). Comparing $\nu_{max}$ for a cloud–fluid medium to our critical value $\nu_{crit}$, we find that

$$\nu_{max} \sim \nu_{crit}. \qquad (23)$$

Using $\nu_{max}$ and equations (10), (13) we find that the minimum viscous inflow time is $T_{vi,min} \approx 100$. We conclude, therefore, that only models with $T_{vi} \gtrsim 100$ (for which $\nu \lesssim \nu_{max}$) may be relevant to real polar rings.

## 3  N–body Simulations

### 3.1  Self–Gravitating Rings

We simulate the evolution of a self-gravitating polar ring using an $N$–body code including an external, rigid, logarithmic potential (equation [1]) to represent the galactic potential. We select units such that $G = M_H = v_o = r_o = 1$ and distribute $N$ particles of total mass $M_R$ (mass ratio $\mu = M_R/M_H$), uniformly within a circular ring of full width $\Delta r$ and full thickness $\Delta z$ centered at $r_o = 1$. We adopt our set of standard model parameters (§2.2) and, in addition, we choose $\Delta z = 0.1$, $N = 30,000$, and $0.03 \leq \mu \leq 0.1$. We give the particles a tangential velocity calculated assuming the orbits are centrifugally balanced according to the radial acceleration found from the net force of the ring and the external potential. We also add a Gaussian distribution of random velocities with tangential ($\sigma_\theta$) and radial ($\sigma_r$) dispersions to avoid axisymmetric instabilities. The adopted Gaussian velocity dispersions obey the relation $\sigma_\theta = \sigma_r \kappa/2\omega_o$ where $\kappa$ is the epicyclic frequency and $\omega_o \equiv v_o/r_o$ is the orbital frequency at $r_o$. We have adopted the value $\sigma_r = 0.1 v_o$ for all simulations. This corresponds to a Toomre stability parameter $Q > 1$ for the initial mass ratios so all rings are initially stable against axisymmetric perturbations. These initial conditions produce only approximate equilibrium states although they are adequate for generating self-gravitating systems for simulations. In practice, after a period of adjustment lasting $\sim 5$ orbits, each self-gravitating ring settles into a smooth configuration that precesses as a whole. The ring is not in equilibrium since mass is continually lost from the inner and outer edges for both numerical (discreteness effects) and physical reasons (see §3.3 below).

We use a tree code (Barnes and Hut 1986; Hernquist 1987; Dubinski 1988) with a leap frog integrator to follow the time–evolution of these rings choosing a timestep $\Delta t = 0.02\tau_o$ corresponding to 50 steps per orbit. We repeated one simulation using $\Delta t = 0.002\tau_o$ to test the accuracy of the integrations. During 30 orbits, the mass, inclination and longitude evolution of the ring track each other within few percent. The Plummer softening radius is 0.03, i.e., approximately 1/3 of the thickness of the ring. The tree code uses an opening angle parameter of $\theta = 1.0$ and quadrupole–order forces in cell–particle interactions. We repeated a simulation with $\theta = 0.7$ and obtained nearly identical results, so we are confident that the adopted parameters are adequate for our purposes. Furthermore, all simulations conserve the total energy to better than 0.2%. The $Z$-component of angular momentum which should be approximately conserved (since the net potential is almost axisymmetric) also varies by no more than 1.0%.

Collisionless simulations of thin disk systems generally require a relatively large number of particles to suppress artificial heating of the disk in the $z$-direction (Sellwood 1987). We use $N = 30,000$ particles for most of the simulations which we believe adequately minimizes discreteness effects. For smaller $N$, a thin ring is quickly heated and breaks apart. For example, a simulation with 3,000 particles only lasts approximately 10 orbits before falling apart while the equivalent simulation with 30,000 particles lasts for a full precession period ($\sim 30$ orbits). We also performed one simulation with 100,000 particles and found good agreement with the equivalent 30,000 particle simulation. Although the smaller simulation loses approximately 20% more mass after one precession period (34 orbits), the two models closely resemble each other for one precession period.

Models with self–gravity are listed in Table 1. Figure 3 presents the evolution of model GB with $\mu = 0.1$ as viewed from the top and the side in the frame coprecessing with the ring. The ring spreads and develops an $m = 2$ spiral mode initially but relaxes to a smooth configuration within 5 orbits precessing as a whole. The $m = 2$ mode is excited by the chosen initial conditions and disappears as soon as orbits mix, allowing the ring to relax to an oval shape. The side view reveals the development of a warp in the ring that curves in the direction of the poles of the potential. An examination of the instantaneous angular momentum of particles as a function of binding energy confirms Sparke's (1991) explanation that the ring warps toward the poles because of inward transport of $Z$-angular momentum. The ring precesses as a solid body and survives for more than a precession period although it eventually breaks up because of continuous mass loss from the inner and outer edges. Nevertheless, self-gravity is sufficient to lock the ring in a state of constant precession lasting at least one precession period. In contrast, the equivalent system of massless test–particles breaks apart because of differential precession within a few orbits.

### 3.2 The Critical Mass Ratio

The simulations listed in Table 1 secure an independent determination of the critical mass ratio $\mu_{crit}$. We find that the minimum mass model that does not break apart from differential precession and survives for one precession period is model GD with initial mass ratio $\mu = 0.045$. However, all rings bleed off mass continually from their inner and outer edges. The escaping mass ends up populating an inner and an outer shell with precessing orbits. This mass loss amounts to approximately half of the mass for each of the models after a full precession period (see Fig. 5 below). This correction for mass loss suggests that rings with an initial mass ratio $\mu \approx 0.04$ and a final mass ratio as low as $\mu = \mu_{crit} \approx 0.02$ have enough self-gravity to lock them into constant precession that lasts more than one precession period. In contrast, model GE ($\mu = 0.03$ initially) does not survive the initial adjustment phase, falling apart within $\sim 5$ orbits. The initial burst of mass loss is enough to take it below the critical mass ratio. Accounting for mass loss, these results are in agreement with our predictions for $\mu_{crit}$ presented in §2.

### 3.3 The Effect of Mass Loss

We determine the mass remaining in the rings as a function of time using the relative inclinations of particle orbits to the midplane of each ring. The direction of the instantaneous angular momentum vector $\mathbf{L}$ of a particle provides a measure of the inclination $i$ of its orbit, i.e., $\cos i = L_z/|\mathbf{L}|$. Orbits with inclinations that differ by more than an angle $|\Delta i_p|$ from the mean inclination of the ring are assumed unbound. We use particles with $|\Delta i_p| < 20°$ to determine the mass of the rings. (We have also used an alternative method in which we sum up the mass inside a torus centered at the mean radius and the mean inclination of the ring. This method yields effectively the same results.) Figure 4 presents the distribution of relative orbital inclinations for different radial bins as a function of time for model GB. The particles that escape from the edges form a roughly uniform distribution in the cosines of the inclinations implying a homogeneously phase-mixed set of precessing orbits.

Figure 5 shows the mass remaining in the ring as a function of time for each of the models. Mass loss amounts to approximately half of the initial mass during the length of the simulations. The rate of mass loss for model GB increases suddenly at the end when the ring begins to break apart. In comparison, the rate of mass loss levels off to a constant value for the lower–mass models GC and GD.

We determine the orientation of the ring as a function of time using an iterative technique. We initially calculate the moment of inertia tensor for the entire set of particles and determine the direction of the principal axes. This provides an initial estimate of the orientation of the ring plane. On successive iterations, we recalculate the inertia tensor using only those particles within a cylindrical volume of full height $0.6\,r_o$ aligned with the particular estimate of the ring plane and rederive a new set of principal axes. Once the number of particles varies by no more than a specified tolerance (100 particles) within the cylindrical volume, we assume we have located the true orientation of the ring plane. Approximately 5 iterations are needed for convergence. The direction of the $z$-component of the inertia tensor then provides both the mean ring inclination and the precessional phase angle. We also varied the radial limits and thickness of the cylindrical volume. Particles with $0.9 < r < 1.0$ are less inclined by $\sim 3°$ than particles with $1.0 < r < 1.1$ revealing the expected warping towards the poles of the potential.

Figure 6 shows the precession frequency $\omega_p$ versus time for the different models in comparison to the predicted frequencies for different inclinations from equation (3). The rings initially precess at a constant rate but at late times the precession frequency begins to decline. The decline in both the mass loss rate and precession frequency results from a secular increase in the inclination of the ring. Figure 7 shows the mean inclination of all the models as a function of time. In all models, the inclination remains fairly constant for many orbits but eventually begins to increase linearly with time. The time of onset of this increase depends on the mass of the ring with the lower–mass models starting first. The increase in inclination lessens the effect of differential precession making it easier for rings to hold on to their mass while also slowing down the precession rate. (The apparent rebound in model GD at $t = 36\tau_o$ is not a real effect. The measured inclination has an intrinsic error of a few degrees since the ring is neither geometrically thin nor perfectly planar.) Figure 8 illustrates the secular increase of the inclination in model GD. Particles lost from the inner and outer edges of the ring carry away $Z$-angular momentum which results in the ring tilting to a higher inclination.

Mass loss is initiated in the models by the adopted initial conditions and is the mechanism that drives the inclination increase. We tested this by removing the shells from simulations GB, GC, and GD. We eliminated all particles inclined by more than $\pm 20°$ from the mean plane of each ring at $t = 10\tau_o$ and continued the simulations. The mean inclination settles to a roughly constant value after a short phase of readjustment in all of these models. A steady inclination increase is no longer observed at late times.

Returning to the original models listed in Table 1, during the initial phase of readjustment ($\approx 5$ orbits), approximately 20% of the mass is lost to inner and outer shells. A comparison of the 30,000 particle model to the 100,000 particle model (Fig. 5) reveals that only about 20% of the mass lost can be attributed to discreteness effects of the $N$-body simulations. The remaining mass loss is intrinsic to the system and may be an important physical effect. Such a process may occur in real polar rings when stars forming from the gas abandon the closed streamlines of gas orbits. Stars that stray beyond the region of stability defined by the self-gravity of the ring may precess away from the plane of the ring. One can imagine that an initial burst of star formation will occur before the ring reaches equilibrium so that mass lost in the form of stars may populate shells surrounding the galaxy.

In light of the above results, the assumption of a *rigid* external potential is questionable. We suggest that future modeling of stellar polar rings should allow for external potentials that respond self-consistently as the embedded rings evolve in time.

### 3.4 *Viscous Rings*

We have introduced the velocity–dependent drag force of equation (21) into the $N$–body code to examine the competition between self–gravity and viscous inflow. We obtain the tangential velocity $v_\theta$ at each time step in the $(r, \theta, z) \equiv (x, y, z)$ frame attached to the ring and we transform the instantaneous values back to the inertial frame $(X, Y, Z)$ of the external potential and the $N$–body code. We use an implicit leapfrog scheme for the integration that requires several iterations at each timestep to assure that the drag and gravitational forces are calculated at the same instant. We tested this method against the more accurate Bulirsch-Stoer method (Press *et al.* 1986) using test–particle integrations and found that the iterative scheme compared well.

Table 2 presents models with $\mu = 0.1$ which include a viscous drag force. The viscosity coefficients and the derived viscous inflow time scales are also listed. The evolution of these models initially resembles that of the dissipationless models although the rings break apart on shorter time scales in response to the viscous forces. Figure 9 shows the evolution of the ring with the largest amount of viscosity (model VD in Table 2). At the time of breakup, the ring splits into several smaller rings that eventually phase–mix to form a shell. The observed breakup times, expressed in orbital periods, are also listed in Table 2 for all models. We determine the breakup time by examining the simulation snapshots. Usually the ring breaks apart within a few orbital periods after our estimated times (Fig. 9). As the evolution proceeds, viscous rings shrink in radius and begin to precess more rapidly which eventually leads to their destruction. This makes sense since $\mu_{crit}$ increases with decreasing radius (equation 16). Figure 10 shows the mean radius of the viscous ring models as a function of time. The mean radius declines steadily with time. According to the analytic model of equation (22), if $r$ is close to $r_o$, the radius should evolve as

$$r = r_o - \frac{\nu}{r_o} t \qquad (24)$$

We see clearly this linear decline of the mean radius in Fig. 10. A linear least squares fit to the lines recovers the values of the input viscosity coefficients implying that the analytic model of §2.4 is a good description of the influence of a velocity–dependent viscous drag. Shrinking together with mass loss drive the ring into a regime where self-gravity can no longer bind individual orbits. Model VD with the shortest viscous time scale breaks up first within half a precession period. (The apparent turn of the mean radius at $t = 24\tau_o$ occurs because the model breaks up.) As the viscous time is increased, the rings break up at later times until the breakup time becomes longer than one precession period and the evolution ends up being very similar to that of the corresponding dissipationless model. The results listed in Table 2 agree with the critical viscous inflow time ($T_{vi,crit} = 850$) derived analytically in §2.4 above.

## 4 Discussion

We have examined the effects of self–gravity and viscous drag on the dynamical evolution of highly inclined rings inside a galactic gravitational potential using analytical and numerical methods. We have stressed the interplay between external gravity, self–gravity, and viscosity while ignoring pressure effects and shock formation in our models. We have chosen appropriate initial conditions: a high inclination of 80° and a full width/radius ratio of 0.4 for the rings and a moderately nonspherical, oblate spheroidal potential with axis ratio $q = 0.85$.

The critical ring/galaxy mass ratio for self–gravitating rings required to avoid destruction through differential precession is low, i.e., $\mu_{crit} \approx 0.02$ (equations [16, 17]). Orbit integrations inside a combined galactic potential and a precessing self–gravitating ring confirm this value for the critical mass ratio (§2.3).

Viscosity cannot be neglected if the viscous spreading time $\tau_{vs}$ (equation [11]) is comparable to the precession period $\tau_p$ (equation [9]), in which case nuclear inflow begins to overpower self–gravity and drives the ring away from its long–lived state of solid–body precession (see also equation [20]). For our moderately extended rings, the critical case occurs when the viscous inflow time $\tau_{vi}$ (equation [10]) is equal to about 25 precession periods (see §2.4). This value is also comparable to the largest possible viscosity expected to be present in galactic matter (see equation [23] and the viscosity estimates of Steiman–Cameron & Durisen 1988).

We have also performed $N$–body simulations of highly inclined self–gravitating rings inside the same galactic potential with and without a viscous drag force (§3) to study the details of the evolution over a long time scale, comparable to the precession period of these rings. The $N$–body simulations confirm the above critical values for the mass ratio $\mu_{crit} \approx 0.02$ and for the viscous inflow time $\tau_{vi,crit} \approx 25\tau_p$. As was expected, rings that lock into solid–body precession establish a warp toward the poles of the potential at larger radii and survive in that state for more than one precession period ($> 30$ orbits). The tendency for self-gravitating collisionless models to warp towards the poles contrasts with the gas–dynamical simulations of Katz & Rix (1992) in which the rings warp away from the poles at larger radii. Indeed, the gaseous and stellar components of the ring in NGC 4650A bend in opposite directions with the stars warping towards the poles and the gas in the outer regions eventually warping away from the poles (van Gorkom *et al.* 1987). The stellar population is relatively young ($\sim 1$ Gyr) based on their blue colors. The new stars on their now collisionless trajectories are probably warping in response to their self-gravity in a similar way to the simulations while the gas has settled through dissipation to a surface of constant precession. Polar rings such as A0136-0801 which show a smooth azimuthal light distribution are probably old systems (Whitmore *et al.* 1987), perhaps as old as their host galaxies. Self-gravity is one possible explanation for the existence of old stellar rings. Our calculations show that only a relatively small mass ratio ($\mu \approx 0.02$ adopting our standard parameters; see §2.2) is required to stabilize a ring against differential precession and this is probably easy to achieve for most observed polar–ring galaxies. Alternatively, stellar rings may be long–lived if the potential is spherical (which may be the case for A0136-0801) or if they were formed from gas that had settled to a preferred orientation in a triaxial potential (Steiman–Cameron & Durisen 1982).

By introducing a velocity–dependent drag force into the $N$–body code that mimics the effect of shear viscosity (§3.4), we confirm that self–gravitating rings are destroyed by continuous nuclear inflow only if the viscous spreading time is comparable to or shorter than the precession period ($\tau_{vs} \lesssim \tau_p$). Otherwise, ring evolution is determined solely by self–gravity and mass loss. If viscosity is important in real polar rings (i. e., if $\tau_{vs} \sim \tau_p$), viscous inflow such as that discussed in this paper is expected to affect the late stages of evolution since it has a noticeable dynamical effect only after $\sim 0.5$–1 precession periods in the evolution of our models.

We have discovered a new physical process which may be important for highly evolved polar rings. Self–gravitating rings suffer extensive mass loss from their edges that in turn drives a secular increase of the mean inclination toward the poles of the potential (Figures 7, 8) after $\sim 10$ orbits. Approximately, half of each ring's mass is lost in the process during a precession period and forms two shells of precessing orbits at the edges of the ring. Only about 20% of the mass loss can be explained by the discreteness of the $N$-body simulations. The remaining 80% of mass is lost during the initial adjustment period and more gradually as a steady leakage from the edges during the evolution of the ring. The additional stellar debris that is observed surrounding some polar–ring galaxies such as NGC 4650A and ESO 415-G26 may be the leftover mass from the original accretion event (Whitmore *et al.* 1987) but could also arise from stars that formed in the gaseous ring and eventually leaked away. Given enough time ($\gtrsim$10 orbits), mass loss drives a secular increase of the inclination. We may therefore expect the stars in older polar rings to be more highly inclined than in younger ones.

## 5  Conclusions

1. *Self-gravity.* A relatively small amount of self-gravity in polar rings (a ring/galaxy mass ratio of $\mu \gtrsim 0.02$) is sufficient to prevent destruction by differential precession. Collisionless self-gravitating rings warp at larger radii towards the poles of the potential as predicted by Sparke (1986). Self-gravity appears to be the most likely stabilizing mechanism for the oldest stellar rings since dissipative forces have no effect on stars. The modest mass required for stability is probably available in most polar–ring galaxies.

2. *Viscous inflow.* Viscous dissipation in polar–ring matter can be anywhere between competitive with precession and negligibly small. If viscosity does compete with precession, it will initiate inflow toward the central host galaxy within just a few ring orbits. If viscosity is negligible, the region between the ring and the galaxy will remain empty and the inner edge of the ring will be sharply defined. Observations of the regions between the central galaxies and their polar rings should provide indirect clues about the true magnitude of viscosity in polar–ring matter.

3. *Mass loss.* The stellar component in polar rings should continually escape from their edges and should form shells of precessing stars. The stellar debris observed around some polar–ring galaxies may be partially explained by this process although some of it is probably left over from the original accretion event.

4. *Inclination increase.* Even if the stars in polar rings do not decouple from the gas by moving to higher inclinations (because most of the mass may be in the gaseous component that is not expected to suffer mass loss), stars in the outer regions will attempt to warp toward the poles of the potential while gas in these regions will warp in the opposite direction (Katz & Rix 1992). Indeed, the gaseous and stellar components of NGC 4650A (van Gorkom *et al.* 1987) generally warp in opposite directions and in the sense expected for each component. A partial decoupling between stars and gas should then be expected in the outer regions of polar rings especially for older systems.

Acknowledgments

We would like to thank Asao Habe, Neal Katz, and Reynier Peletier for many stimulating discussions and Linda Sparke and the referee, Thomas Steiman–Cameron, for many comments and suggestions. DMC was supported in part by NSF PYI award AST 91–48279 (R. Narayan) and by NASA grant NAGW–1510 (S. Balbus & J. Hawley). JD is supported by a CfA Postdoctoral fellowship.

Table 1
Self-Gravitating Rings
$q = 0.85$, $i_o = 80°$, $N = 30,000$

| Model | $\mu$ |
|---|---|
| GA[a] | 0.100 |
| GB | 0.100 |
| GC | 0.060 |
| GD | 0.045 |
| GE | 0.030 |

[a] $N = 10^5$

Table 2
Viscous Rings
$\mu = 0.10$, $q = 0.85$, $i_o = 80°$, $N = 30,000$

| Model | $\nu/r_o v_o$ | $T_{vi}$ | $T_{breakup}$ |
|---|---|---|---|
| VA | $1.3 \times 10^{-4}$ | 1200 | ... |
| VB | $2.0 \times 10^{-4}$ | 800 | 31 |
| VC | $4.0 \times 10^{-4}$ | 400 | 27 |
| VD | $8.0 \times 10^{-4}$ | 200 | 20 |

# FIGURE CAPTIONS

FIG. 1.— Orbits starting at $r = r_-$ (left) and at $r = r_o$ (right) are projected in three different planes for more than two precession periods (up to $t = 80\tau_o$). The precessing ring sits on the $xy$ plane and has a mass ratio $\mu = 0.03$ above the critical value. Both orbits continue to precess together with the self–gravitating ring indicating they are parts of a long–lived configuration established by the ring's self–gravity.

FIG. 2.— As in Figure 1, but for a mass ratio $\mu = 0.01$ below the critical value. Both orbits precess away from the $xy$ plane of the precessing ring.

FIG. 3.— The evolution of a self-gravitating ring with initial $\mu = 0.1$ (model GB) in a frame precessing with the ring. A top view and a side view are shown for each time. The symmetry axis of the oblate spheroidal potential is parallel to the vertical axis in the side view. The times are in units of orbital periods, $t_o$. After a period of adjustment lasting $\sim 5$ orbits, the ring precesses as a solid body for more than 30 orbits. As predicted by Sparke (1986), the ring warps toward the poles of the potential. In addition, the ring suffers from mass loss that populates an inner and an outer shell of precessing orbits.

FIG. 4.— The distribution of cosines of particle inclinations relative to the midplane of the ring versus time for model GB. The times are in units of orbital periods, $t_o$. Most particles remain within the ring but particles from the inner and the outer edge form shells indicated by the uniform distribution in $\cos i$.

FIG. 5.— Mass remaining in the self-gravitating rings versus time. The models lose approximately half of their mass during their evolution. The initial high rate of mass loss occurs while the rings adjust to a smooth configuration.

FIG. 6.— Precession frequencies versus time for the self-gravitating ring simulations. The horizontal lines denote the precession frequencies predicted by equation (3) with $r = r_o$ at different inclinations as labeled. The precession frequencies of the rings remain constant and close to the predicted value for $i_o = 80°$ but decline at late times in response to a secular increase in the inclination of the ring.

FIG. 7.— The mean inclination of self-gravitating rings versus time. The inclination fluctuates by only 2–3 degrees from the initial value for many orbits but eventually begins to increase linearly with time. This secular change of the inclination is driven by the formation of mass shells that remove $Z$-angular momentum from the ring.

FIG. 8.— The evolution of model GD ($\mu = 0.045$) viewed edge on in the precessing frame. As the ring loses mass, the inclination increases and the precession frequency declines demonstrating a net loss of $Z$-angular momentum from the ring to the shells of escaping particles.

FIG. 9.— Time–evolution of a viscous self–gravitating ring with $\mu = 0.1$ and $T_{vi} = 200$ (Model VD in Table 2). The times are in units of orbital periods, $t_o$. Initially the ring evolves as model GB, although it gradually flows inward. After about 20 orbits, the ring breaks into several ringlets and eventually phase–mixes within a shell.

Fig. 10.– Time–evolution of the mean radius for the viscous ring models. The mean radius shrinks at the linear rate predicted by equation (24).